\RequirePackage{lineno}
\documentclass[aps,prl,twocolumn,showpacs]{revtex4}
\usepackage{amsmath,amssymb}
\usepackage{graphicx}
\usepackage{amsfonts}

\def\huga#1{\begin{gather} #1 \end{gather}}

\def\hual#1{\begin{align} #1 \end{align}}

\newcommand{\R}{{\mathbb R}}

\def\ga{\gamma}\def\th{\theta}

\def\pa{{\partial}}
\newcommand{\bi}{\begin{itemize}}\newcommand{\ei}{\end{itemize}}
\newcommand{\ben}{\begin{enumerate}}\newcommand{\een}{\end{enumerate}}
\newcommand{\bce}{\begin{center}}\newcommand{\ece}{\end{center}}
\newcommand{\reff}[1]{(\ref{#1})}

\newcommand{\hs}[1]{{\hspace{#1}}}\newcommand{\vs}[1]{{\vspace{#1}}}

\def\eps{\varepsilon}
\def\ra{\rightarrow}

\newcommand{\barr}{\begin{array}}\newcommand{\earr}{\end{array}}
\newcommand{\bpm}{\begin{pmatrix}}\newcommand{\epm}{\end{pmatrix}}
\newcommand{\bsm}{\left(\begin{smallmatrix}}
\newcommand{\esm}{\end{smallmatrix}\right)}
\newcommand{\ba}{\begin{array}}\newcommand{\ea}{\end{array}}

\def\Om{\Omega}

\def\del{\delta}

\def\ig{\includegraphics}
\begin{document}
%\setpagewiselinenumbers\modulolinenumbers[5]\linenumbers
\title{A New Type of Traveling Interface Modulations in a 
Catalytic Surface Reaction}
\author{M. Rafti} \affiliation{Instituto de Investigaciones Fisicoqu\'imicas
Te\'oricas y Aplicadas (INIFTA), Fac. Cs. Exactas, Universidad Nacional 
de La Plata, Calle 64 y diag. 113 (1900), La Plata, Argentina}
\author{H. Uecker}  \affiliation{ Institut f\"ur Mathematik, Carl von 
Ossietzky Universit\"at Oldenburg, D-26111 Oldenburg, Germany} 
\author{F. Lovis} \affiliation{Institut f\"ur Physikalische Chemie und 
Elektrochemie, Universit\"at Hannover, Callinstr. 3 - 3a, D-30167 Hannover, 
Germany}
\author{V. Krupennikova}\affiliation{Institut f\"ur Physikalische Chemie und 
Elektrochemie, Universit\"at Hannover, Callinstr. 3 - 3a, D-30167 Hannover, 
Germany}
\author{R. Imbihl}\affiliation{Institut f\"ur Physikalische Chemie und 
Elektrochemie, Universit\"at Hannover, Callinstr. 3 - 3a, D-30167 Hannover, 
Germany}
\begin{abstract} 
A new type of traveling interface modulations has been observed in the 
NH$_3$ + O$_2$ reaction on a Rh(110) surface. A model is set up 
which reproduces the effect, which is attributed to diffusional mixing of two 
spatially separated adsorbates causing an excitability which is strictly 
localized to the vicinity of the interface of the adsorbate domains. 
\end{abstract}
\maketitle
%\section{Introduction}
Pattern formation in reaction-diffusion systems covers a wide range of 
fascinating phenomena in liquid phase chemistry, biochemistry, biology and 
catalytic surfaces \citep{ch93,ks95,ertl-i95}. In general, the 
patterns arise due to the coupling of a non-linear reaction term 
with diffusion. Reaction fronts, target patterns and 
spiral waves, stationary concentration patterns and 
chemical turbulence have been seen. Various additional factors like global 
coupling, diffusional anisotropy, energetic interactions and cross 
diffusion of reactants may add to the complexity and diversity of the chemical 
wave patterns. 

Extended bistable systems generically exhibit  
fronts (also called interfaces or domain walls) 
connecting one phase in one part of the spatial domain 
to the other phase in some other part of the domain. 
In two spatial dimensions the most natural geometry is a straight 
line for the front position, suitably defined as some intermediate level curve 
of the solution. However, already in simple bistable systems, initially straight 
interfaces between two domains may undergo 
a number of instabilities, see, e.g., \cite[Chapter 2]{pis06} for 
an overview. 
A typical case is a linear transverse instability leading 
to a regular (periodic) or irregular bending of the 
front, but with small amplitude, which may then often be described 
by Kuramoto-Sivashinksky type of equations, see \cite{kur84}. 
Another possibility is that an instability does not saturate at some small 
amplitude, which may yield ``fingering'' and 
labyrinthine patterns \cite{LMO93,GMP96,HYY06}. 
Similar wave instabilities also occur in excitable media, see, e.g., 
\cite{ZMM98}. 

Here we report on a new type of instability and self-organization
of an interface, namely 
interface modulations that originate from corners and travel 
along the interface in a pulse like fashion, leaving the interface 
position almost unperturbed behind. 
These excitations have been observed in the NH$_3$ + O$_2$ 
reaction on a Rh(110) surface. The effect is attributed to diffusional 
mixing of two spatially separated adsorbates causing an excitability which
 is strictly localized to the vicinity of the interface of the adsorbate 
domains. 
Combining a bistable with an excitable system, we set up a general model 
which reproduces the traveling interface modulations seen in the experiment.

%\section{Experimental results}
\bigskip%\vs{4mm}
The reaction we study is the catalytic ammonia oxidation with O$_2$ on a 
Rh(110) surface under low pressure 
conditions ($10^{-5}$ mbar) in a UHV chamber 
equipped with a photoemission electron microscope (PEEM) as spatially 
resolving method.  Illuminated with a D$_2$ discharge lamp (5.5--6 eV) 
photoelectrons are ejected which allow an imaging of the local work 
function with a spatial resolution of $\approx 1\mu$m and the temporal 
resolution of video images (20 ms).  At elevated temperatures 
(T$>$400 K) both reactants dissociate upon adsorption 
into their atomic constituents O$_{ad}$, N$_{ad}$, and H$_{ad}$ 
\cite{cdk98,kbr95}. 
The atomic adsorbates recombine, forming N$_2$, NO, and H$_2$O as main 
products. Also, H$_2$ is produced and desorbs at a high rate, and hence 
the coverage $\th_H$ is always small. The adsorbates N and O form a large number 
of ordered reconstruction phases on Rh(110) but under our 
reaction conditions only the (2$\times$1)-N/(3$\times$1)-N corresponding  
to $\th_N=0.5/0.33$ and  the c(2$\times$6)-O corresponding 
to $\th_0=0.66$ are relevant \cite{gmo95}.  
In addition, a mixed coadsorbate phase c(2$\times$4)-2O,N may form. 
\begin{figure}[htbp]
\bce 
\ig[width=80mm]{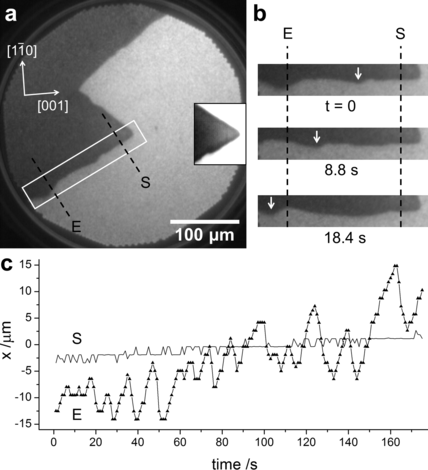}
\ece
\caption{{\small Experimental observation of interface excitations in the 
NH$_3$ + O$_2$ reaction on Rh(110). Experimental conditions: 
T=740 K, p(NH$_3$)=$3.85\times10^{-5}$mbar, 
p(O$_2$)=$1.35\times10^{-5}$ mbar. 
(a) PEEM image showing the interface between oxygen covered (dark) 
and nitrogen covered surface area (bright). The inset representing 
an enlarged view of the interface region near S shows the formation 
of dark boundary layer at the interface within the oxygen phase.
(b) Enlarged view of the region marked in (a) showing the pulse-like 
propagation of an interface modulation.
(c) Position vs.~time plots showing the temporal variation of the 
interface position. The data were taken from cross sections perpendicular 
to the interface at points E and S in (a).
 \label{f0}}}
\end{figure} 

Over a broad range of parameters the reaction exhibits simple bistability, 
i.e.~one observes a broad hysteresis in the reaction rates in 
heating/cooling cycles. The reactive branch is associated with the 
(2$\times$1)/(3$\times$1) of nitrogen, 
the unreactive branch with the c(2$\times$6) of oxygen. 
Transitions between the two states occur via fronts. If one adjusts 
conditions close to equistability both phases are simultaneously 
present as shown by the PEEM image in Fig.1a. 
Since oxygen adsorption strongly increases the work
function (WF) ($\Delta\Phi_{\max}\sim 1.0\,$eV) 
high O$_{ad}$ coverages are imaged dark 
whereas adsorbed nitrogen which only 
causes a maximum WF increase of 
280 meV appears bright \cite{MSI97}. 

The position of the interface is nearly stationary but one notices
small lateral displacements of the interface which emanate near the
sharp corner in the phase boundary and then propagate in a pulse-like
manner along the interface. This process is depicted in more detail by
the frames in Fig.1b displaying an enlarged section of the PEEM image
in (a).  The velocity of the pulse-like excitations is about 
6 $\mu$m/s. Cross sections of the interface showing the temporal variation
of the interface positions at two different points, E and S, are
displayed in Fig.1c. Near the sharp corner, in point S, the amplitude
is below the detection limit. Further away, at point
E, the amplitude is substantial varying between a few $\mu$m 
and 20 $\mu$m. One notes a drift of the average interface 
position of about 15 $\mu$m over
an observation time of 170 s which is due the fact that the
equistability conditions are not exactly met. 
The time series exhibits irregular behavior but the excitability 
of the interface is quite stable and can be observed over 
several hours. The average period of the local 
excitations is around 10 s. In our experiments we
found no correlation between the interface angles and 
interface excitations, and the
crystallographic directions of the surface.  

In order to understand why the
excitations remain localized at the interface and do not extend into the
interior of the phase it is helpful to look into the chemically
rather similar system Rh(110)/NO + H$_2$ which can be considered as well
understood \cite{MInat94,MSI97}.  
Some spectacular chemical wave patterns including
rectangularly shaped target patterns and spiral waves and traveling
wave fragments were found there. The excitable behavior in this system
was shown to be based on a cyclic change of three different
structures; the c(2$\times$6)-O of oxygen, the 
(2$\times$1)/(3$\times$1)-N of nitrogen and
the c(2$\times$4)-2O,N as mixed coadsorbate phase. In the NH$_3$ + O$_2$ reaction
only two of these three structures are present as stable phases while
the mixed coadsorbate phase is missing. Apparently the mixed phase
does not form by coadsorption.  

If we assume that by surface diffusion
this mixed phase may form, its formation is restricted to a
boundary layer along the interface where the two separated adsorbates,
N and O, can penetrate each other by diffusion. Excitability would
then be strictly restricted to a boundary region along the interface
and this is what we basically see in the experiment. Using the
diffusion values which have been used for quantitative simulation of
the chemical wave patterns in Rh(110)/NO+H$_2$ we can estimate the
diffusion length {\it l} at T=740 K for $\tau$=10 s with ${\it l}=\sqrt{2 D\tau}$ 
resulting in ${\it l}= 8\,\mu$m for N and 13$\mu$m for O \cite{MSI97}.  
The inset in Fig.~1a shows a dark boundary 
region of a few $\mu$m width which is consistent with the
high WF of 1.1 eV of the c(2$\times$4)-2O,N phase \cite{MeSI97}.

%\section{A general model}
\bigskip
For modeling the observed behavior we set up a dimensionless 
3-variable model for 
bistable/excitable media which in 2D reads
\begin{subequations}\label{ps2} 
\hual{\pa_t u=&u-u^3-v-\del(u{-}u_s)q^2+d_u\Delta u+d_{uq}\Delta q, \\
\pa_t v=&\eps(u+\beta-v)+d_v\Delta v,\\
\pa_t q=&(1-q)(q-a)(q+1)+\ga(1-q^2)(u-u_s)\notag\\
&+d_{uq}\Delta u +\Delta q, 
}
\end{subequations}
with diffusion constants, $d_u,d_{uq},d_v>0$, parameters 
$\beta,\ga,\del\in\R$, $\eps>0$ and $-1<a<1$. 
In short, using  $U=(u,v,q)$ with obvious notations we write 
\huga{\label{ps2b}
\pa_t U=f(U)+D\Delta U.
}
The system \reff{ps2} is composed of an excitable 
$u,v$--subsystem (FHN like) 
and a bistable $q$--subsystem (Allen-Kahn or Nagumo equation), 
which has front solutions. 

The basic idea is that (i) through the 
interaction with the $q$-variable the $u,v$-subsystem is excitable 
only in the vicinity of the front position (where $q\approx 0$), 
and that (ii) these localized 
excitations of the $u,v$--subsystem then push or pull the $q$-front. 
Since on surfaces the diffusion of the different species is not 
independent of each other, we include cross-diffusional terms 
which have to be symmetric according to Onsager’s reciprocity relation. 
On surfaces cross diffusion arises (i) due to the vacant site requirement
 for diffusional hops and (ii) due to energetic interactions between 
coadsorbed species \cite{ELT02,ve09}. In particular, the strong repulsive 
interaction between 
coadsorbed oxygen and nitrogen shows up in a downward shift in the 
$N_2$ desorption maximum by about 100 K \cite{Com92}. 
As will be 
shown below cross-diffusion becomes important for the nucleation of 
excitation pulses.   

Thus, we choose parameters $\beta,\eps$ in such a way 
that in the absence of $q$, i.e., for $q\equiv 0$, 
the $(u,v)$ ODE subsystem $\pa_t (u,v)=(f_1(u,v,0), f_2(u,v,0))$ 
is excitable. 
Its unique ODE fixed-point $(u_s,v_s)$ is given by 
$u_s=-\beta^{1/3}$, $v_s=u_s+\beta$. This fixed point 
is asymptotically stable and globally attracting, but for small 
$\eps>0$ rather 
small perturbations may lead to large excursions. 

For $u\equiv u_s$, or equivalently $\ga=d_{uq}=0$, (\ref{ps2}c) is a 
standard bistable equation 
\huga{\label{aco}
\pa_t q=g(q)+\Delta q,\quad g(q)=(1-q)(q-a)(q+1), 
}
i.e., the kinetics $\pa_t q=g(q)$ has the two stable fixed points 
$u=\pm 1$ and the unstable fixed point $q=a$. It is well known, 
that (\ref{ps2}c) has travelling front solutions, e.g., 
$q(x,y,t)=q_f(x-c_0t)$, $q_f(\xi)\ra\pm 1$ as $\xi\ra\pm\infty$, 
in fact explicitly given by 
$
c_0=\sqrt{2}a\text{ and }q_f(\xi)=\tanh(\xi/\sqrt{2}). 
$
For $a<0$ ($a>0$) fronts travel left (right), meaning that the 
$+1$ phase invades the $-1$ phase (resp.~vice versa). 

\begin{figure}[htbp]
\bce 
\ig[width=40mm, height=35mm]{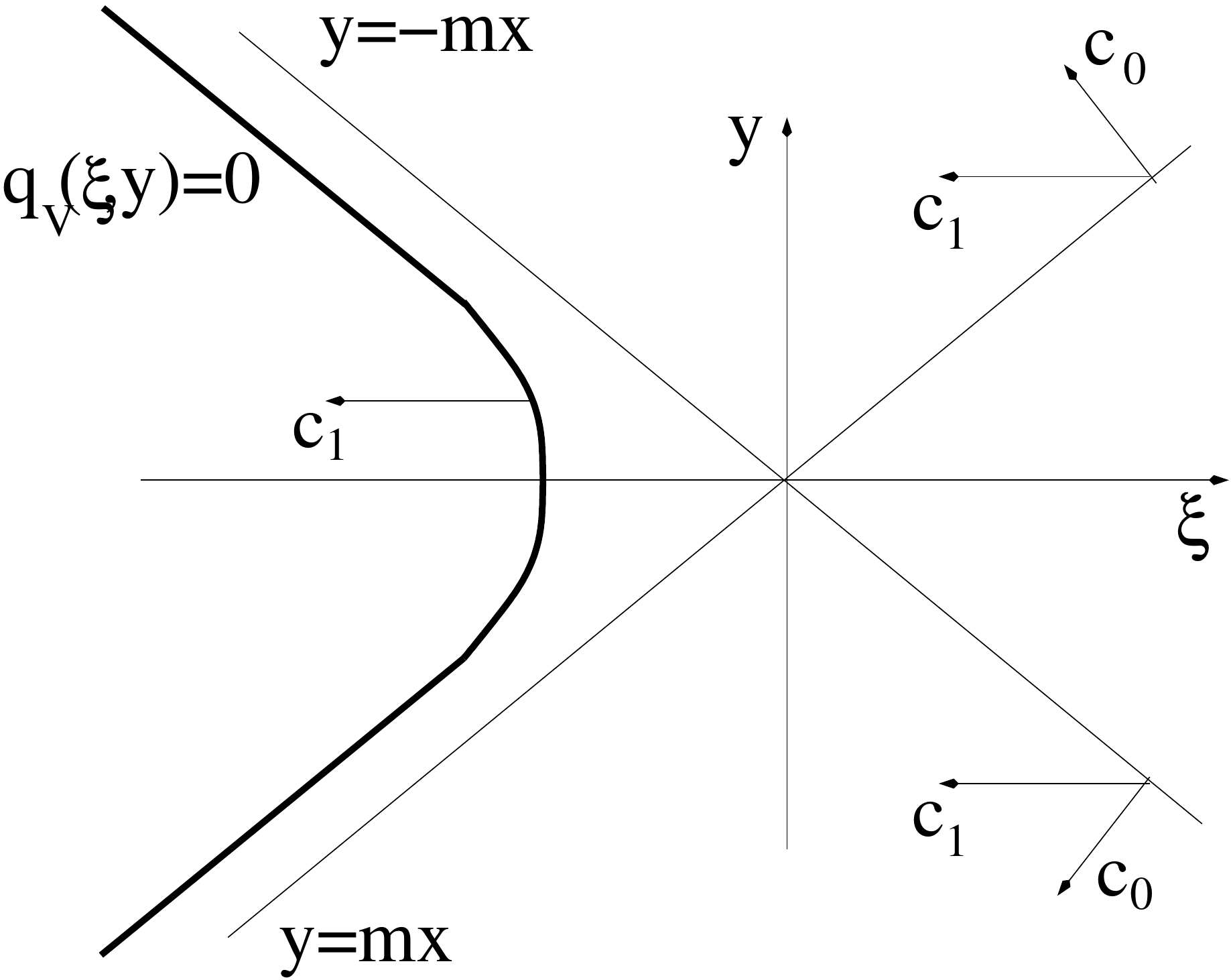}
\ece

\vs{-5mm}
\caption{{\small Heuristics for V-shaped fronts of \reff{aco}. 
\label{fwf}}}
\end{figure}

Since the Laplacian is isotropic 
any orientation of fronts is allowed. 
As a consequence, \reff{aco} 
also has (smooth) V--shaped fronts $q_V$, propagating with 
speed $c_1=c_0\sqrt{1+1/m^2}$, see Fig.~\ref{fwf} and \cite{nt05}. 

Now considering the coupling between (\ref{ps2}a,\ref{ps2}b) and 
(\ref{ps2}c) in more detail we note that $|d_{uq}\Delta q|$ becomes large 
near corners of the front, and vanishes away from the front; 
thus $(u,v)$ excitations originate near corners. On the other 
hand, the term $-\del(u-u_s)q^2$ in (\ref{ps2}a) makes the 
$(u,v)$ kinetics less 
excitable away from the front, 
and thus excitations in the PDE \reff{ps2} stay near the front. 
% \begin{figure}[htbp]
% \bce 
% \ig[width=35mm]{ex0}\quad\ig[width=35mm]{ex1}
% \ece
%
% \vs{-5mm}
% \caption{{\small 
% The $(u,v)$ ODE system for $\beta=0.2, \eps=0.03$, $\del=0.5$ 
% and $q=0$ (easily excited, left) 
% resp.~$q=1$ (larger perturbation needed for excitation, right). \label{f1}}}
% \end{figure} 
Finally, the term $\ga(1-q^2)(u-u_s)$ in 
 (\ref{ps2}c) has the effect that the excitations push or pull 
the $q$--front, as seen in the experiment. 

System \reff{ps2} was integrated numerically  in a domain $\Om=[-L,L]^2$ 
for various parameters 
using different initial conditions (IC) $(u,v,q)|_{t=0}=(u_0,v_0,q_0)$ 
and boundary conditions (BC). 
For the IC we are led by the experiment to consider ``wedges'' in $q$, 
e.g.
\huga{\label{nic}
q_0(x,y)=\left\{\barr{ll} -1&x<x_0-m|y|\\
1&x\ge x_0-m|y|\earr\right., 
}
where $\pm m\in \R$ are the slopes of the sides and $x_0{\in}\R$ 
represents the position of the tip. For $(u,v)$ we choose 
the fixed point $(u_0,v_0){=}(u_s,v_s)$. 
Given an IC of the form \reff{nic}, it is natural to integrate 
\reff{ps2b} in a moving frame $\xi=x-\eta t$ with $\eta\approx c_1(m)$ 
to keep the tip of the wedge away from boundaries, i.e., to 
integrate 
\huga{\label{ps2t}
\pa_t U=f(U)+D\Delta U+\eta\pa_\xi U.
}
For the BC the problem then is that 
while planar fronts can be easily simulated with Neumann BC, 
for V--shaped fronts influences of boundaries on the fronts are difficult 
to avoid. 
Here we choose  Dirichlet BC for \reff{ps2t}, namely 
\hual{\label{nbc}
&\begin{split}
&(u,v)|_{\pa\Om}=(u_s,v_s)\text{ and }
q=\pm 1 \text{ on } \xi=\pm L, \\
&\text{ $q(\xi,\pm L)=q_f(\xi-\xi_0)$}. 
\end{split}
}
The latter fixes the front shape and position 
at the top and bottom boundary. 

\begin{figure}[htbp]
\bce 
\ig[width=41mm]{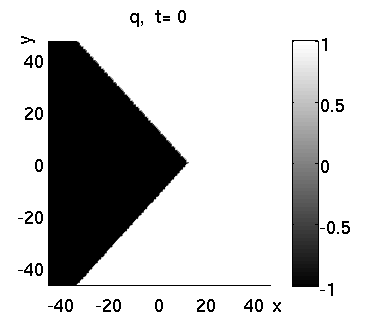}
\ig[width=41mm]{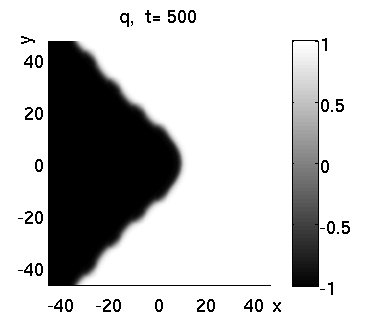}\\
\ig[width=41mm]{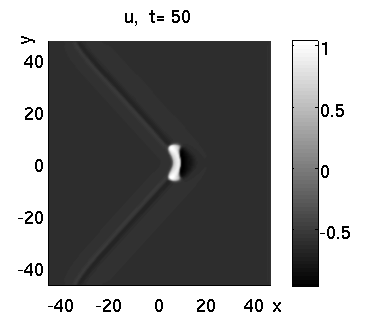}
\ig[width=41mm]{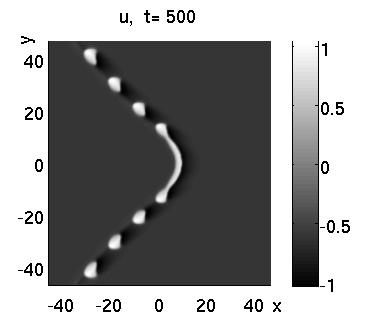}
\ece
\caption{{\small Numerical integration of \reff{ps2} in frame moving with 
speed $\eta=3c_1/4=-0.15$.  
Parameters $d_u=0.09, d_v=0.01, d_{uq}=0.1, 
\beta=0.2, \del=0.5, \eps=0.03, \ga=-0.05, a=-0.1$. 
IC for $q$ is the wedge \reff{nic} with $x_0=L/4$, $m=1$, 
ICs for $(u,v)$ are $(u_s,v_s)$.  BC according to \reff{nbc} with 
$\xi_0=-3L/4$. 
\label{f3}}}
\end{figure}

For the IC and BC chosen above, we obtain the simulation results displayed 
in Fig.~\ref{f3}. Excitations nucleate at the tip of the wedge 
and then travel along the front, pushing it back and forth. 
The chosen $\ga=-0.05<0$ means that 
$u>u_s$ ($u<u_s$) pushes $q$ down (up), such that here 
the excitations push back the frontline. 
The firing process at the tip repeats 
for some time (essentially depending on the size of the computational 
domain), and the process is accompanied  
by some overall reshaping of the wedge. 
Aside from boundary effects, 
this reshaping is determined by the following factors. 
The $q$-front does not fully recover its former position 
 after a $(u,v)$ pulse has passed. 
The tip of the wedge, where pulses 
nucleate, drifts to the right. To counteract this effect we 
chose $\eta=3c_1/4$ (instead of $\eta=c_1$ which without 
coupling to the $(u,v)$ system would give a stationary 
tip position). As a consequence of decreasing $|\eta|$, the unperturbed 
sides of the wedge drift to the left.  
The overall balance gives an almost stationary average 
front position up to $t=500$. 
For $t>500$ excitations that have 
emanated from the tip are reflected by the boundary.  
 
\begin{figure}[htbp]
{\small a) $\ga=-0.2$\hs{25mm} b) $\ga=0.1$}\\
\ig[width=43mm]{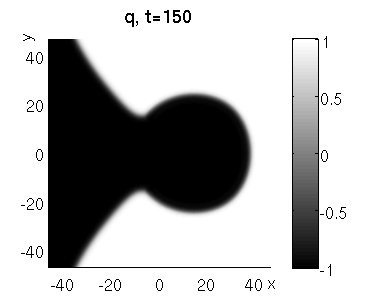}\ig[width=43mm]{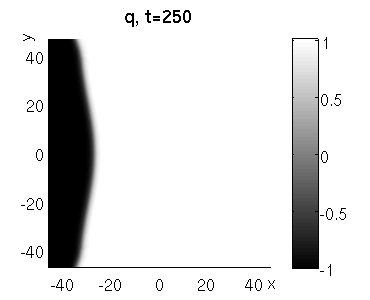}
\caption{{\small Same parameters and IC as in Fig.\ref{f3} 
except for $\ga$. 
\label{f5}}}
\end{figure}

The behaviour in Fig.~\ref{f3} is quite robust with respect to most 
parameters and 
IC’s, including the opening angle of the wedge. 
A decisive parameter for the evolution is $\ga$. For $\ga=-0.2$ 
the excitations push the front too strongly  thus destroying the wedge 
by creating a bubble.  
For $\ga=0.1$ the excitations pull the front too strongly thus flattening 
the wedge, see Fig.~\ref{f5}.

%\section{Summary} 
\bigskip 
In summary, we observed excitability
in a catalytic surface reaction     which
remained strictly localized at the interface of two domains of
different adsorbates. The excitations were traveling along
the interface in a pulse--like way, 
causing lateral displacements of the interface
position. Mechanistically, the localized excitability can be traced
back to the diffusive mixing of the two separate adsorbates at the
interface causing the formation of a mixed coadsorbate phase which is
required to make the system excitable. The experimentally observed
behavior could be reproduced with a general dimensionless 
3-variable model which couples
the excitability of a subsystem to the position of a frontline.  The
nucleation of excitations at corners of the front was explained with
cross-diffusional effects which are very sensitive to the local front
geometry. Similar dynamical behavior should be expected in all systems
which (i) are essentially bistable in the sense that 
there are two asymptotically stable phases, 
but where (ii) diffusive mixing at the interface can locally 
change the dynamical behavior from bistable to excitable.

%\bibliography{./iex}
%\input{iex7.mod.bbl}

\end{document}